%% file: sample-sigconf.tex
\documentclass[sigconf,authorversion]{acmart}

\AtBeginDocument{%
  }
\acmConference[CF '24 Companion]{Proceedings of the 21st ACM International
Conference on Computing Frontiers Workshops and Special Sessions}{May 7--9,
2024}{Ischia, Italy}
\acmDOI{10.1145/3637543.3652873}

\begin{document}
\title[Integrating SystemC-AMS]{Integrating SystemC-AMS Power Modeling with a RISC-V ISS for Virtual Prototyping of Battery-operated Embedded Devices}
\author{Mohamed Amine Hamdi$^1$, Giovanni Pollo$^1$, Matteo Risso$^1$, Germain Haugou$^2$, Alessio Burrello$^1$, Enrico Macii$^1$, Massimo Poncino$^1$, Sara Vinco$^1$ and Daniele Jahier Pagliari$^1$}

\affiliation{%
  \institution{$^1$Politecnico di Torino, Turin, Italy - first\_name.first\_surname@polito.it;} \city{}  \country{}
}
\affiliation{%
  \institution{$^2$GreenWaves Technologies SAS, Grenoble, France - germain.haugou@greenwaves-technologies.com} \city{}  \country{}
}

\renewcommand{\shortauthors}{Hamdi, Pollo et al.}

\begin{abstract}
\input{sec/abstract}
\end{abstract}

\ccsdesc[500]{Hardware~Power estimation and optimization}
\ccsdesc[500]{Hardware~Simulation and emulation}

\keywords{SystemC, RISC-V, Virtual Prototyping, Power, Design Space Exploration}

\maketitle
\section{Introduction}\label{sec:intro}
\input{sec/introduction}

\section{Virtual Prototyping Architecture}\label{sec:methods}
\input{sec/methods}

\section{GVSoC-SystemC-AMS Integration}\label{sec:integration}

\input{sec/integration}

\vspace{-0.2cm}
\section{Experimental Results}\label{sec:results}

\input{sec/results}

\section{Conclusions}\label{sec:conclusions}
\input{sec/conclusions}

\begin{acks}
This work has received funding from the Key Digital Technologies Joint Undertaking (KDT-JU) under grant agreement No 101095947 and grant agreement No 101112274. The JU re-ceives support from the European Union’s Horizon Europe research and innovation programme.
This publication is part of the project PNRR-NGEU which has received funding from the MUR – DM 117/2023.
\end{acks}

\bibliographystyle{ACM-Reference-Format}
\bibliography{bibliography}

\end{document}

%% file: sec/abstract.tex
RISC-V cores have gained a lot of popularity over the last few years.
However, being quite a recent and novel technology,  there is still a gap in the availability of comprehensive simulation frameworks for RISC-V that cover both the functional and extra-functional aspects. This gap hinders progress in the field, as fast yet accurate system-level simulation is crucial for Design Space Exploration (DSE).

This work presents an open-source framework designed to tackle this challenge, integrating functional RISC-V simulation (achieved with GVSoC) with SystemC-AMS (used to model extra-functional aspects, in detail power storage and distribution).
The combination of GVSoC and SystemC-AMS in a single simulation framework allows to perform a DSE that is dependent on the mutual impact between functional and extra-functional aspects.
In our experiments, we validate the framework's effectiveness by creating a virtual prototype of a compact, battery-powered embedded system.

%% file: sec/introduction.tex
RISC-V has become a pivotal player in the semiconductor industry by offering a customizable and free alternative to the conventional, proprietary ISAs \cite{riscv}. Its widespread adoption in various sectors, from embedded systems to high-performance computing, highlights its potential to transform the design and implementation of computing systems. 

The interest of the research community led to the development of many frameworks and implementations supporting the functional simulation and customization of RISC-V cores \cite{riscv_simulation}. The goal of such frameworks is to allow the development of virtual prototypes, enabling designers and engineers to model, analyze, and optimize systems based on RISC-V before constructing physical prototypes \cite{riscv-vp,8524047}. This drastically cuts down on development time and costs, allows for the early identification of design issues, and supports the exploration of various architectural configurations to achieve specific performance, power, and area objectives.

However, all such solutions focus on functionality and timing-related aspects, thus allowing only a partial DSE. This is a severe limitation: the high degree of heterogeneity, the technology challenges, and the tight coupling with the environment of modern systems require indeed to \emph{also consider extra-functional metrics} such as power consumption and thermal behavior to ensure correct operations \cite{7812663}. It is thus necessary to \emph{monitor the evolution over time of functionality not in isolation, but rather together with extra-functional properties}. As highlighted in \cite{riscv-survey}, RISC-V simulators provide at most an analysis of the total energy consumption, %
with no integrated simulation of functionality and power dynamics. 

Outside of the RISC-V community, SystemC and its Analog Mixed Signal (AMS) extension \cite{systemc-ams} emerged as a widespread solution for the creation of virtual platforms \cite{8351864}. SystemC-AMS facilitates high-level modeling and simulation of mixed-signal systems, supporting a wide range of components. Additionally, it can effectively cover extra-functional behaviors, ranging from mechanical systems to power dynamics \cite{7812663,9969885}. Thus, SystemC-AMS seems a viable solution for the construction of power-aware virtual platforms.    

This paper proposes an open-source framework that integrates SystemC-AMS with GVSoC~\cite{gvsoc}, a RISC-V Instruction Set Simulator (ISS), to build an open-source virtual platform that is aware of both functionality and extra-functional aspects, with a focus on power. The goal is to allow the simulation of the power flows in the system, not only in terms of power demand of the RISC-V core but also considering, for instance, the battery subsystem that feeds the functional parts. With experiments on a simulation setup that models a simple battery-operated embedded device for audio processing, we show that our framework enables detailed and extensive DSE, with an acceptable simulation time overhead with respect to vanilla GVSoC\footnote{The source code of our simulator is available at \href{https://github.com/eml-eda/messy}{https://github.com/eml-eda/messy}}.

%% file: sec/methods.tex
\subsection{Power Simulation with SystemC-AMS}
SystemC-AMS is an extension of SystemC that allows modeling the analog-mixed signal part of an embedded system.
It covers a wide range of domains thanks to three different modeling styles.  
{Timed Data Flow (TDF)} relies on static scheduling and is used to represent discrete-time systems. {Linear Signal Flow (LSF)} adopts linear algebraic equations to model control system modeling. {Electrical Linear Networks (ELN)} offers a collection of standard linear electrical components, like resistors and capacitors.
When building the simulatable implementation of a component, one can determine the suitable SystemC-AMS flavor by considering the desired (i) level of abstraction, i.e., discrete time or continuous time, (ii)  level of detail of the description (e.g., based on equations or transfer functions), and (iii) adherence to conservation laws. 

SystemC-AMS simulation is event-driven, where a centralized scheduler controls the execution of processes
based on events, i.e., synchronizations, time notifications,
or signal value changes.

The suitability of SystemC-AMS for the simulation of power systems has been investigated in the literature, ranging from small battery-powered systems \cite{10.1145/2627369.2627657} to electric vehicles \cite{8648435}. Power components may be described at a functional level (e.g., as an efficiency equation in case of DC-DC converters) or as a circuit model (e.g., in case of a battery). This implies a corresponding choice of SystemC-AMS constructs, i.e., TDF in the former case and ELN in the latter. 

In SystemC-AMS simulations, the power flow is simulated and managed by a \emph{power bus}, that extends the physical Charge Transfer Interconnect bus (CTI) with the simulation of the power flow, i.e., the aggregation of load demands to determine the amount of current required from the energy sources and storage devices~\cite{9969885}. All components are connected (possibly via DC-DC converters) to the power bus, which has a role very similar to a functional bus in managing the overall power flow (rather than information flow).%

\subsection{Proposed architecture and requirements}\label{subsec:arch}
\begin{figure}[th]
\centering
\includegraphics[width=.5\linewidth]{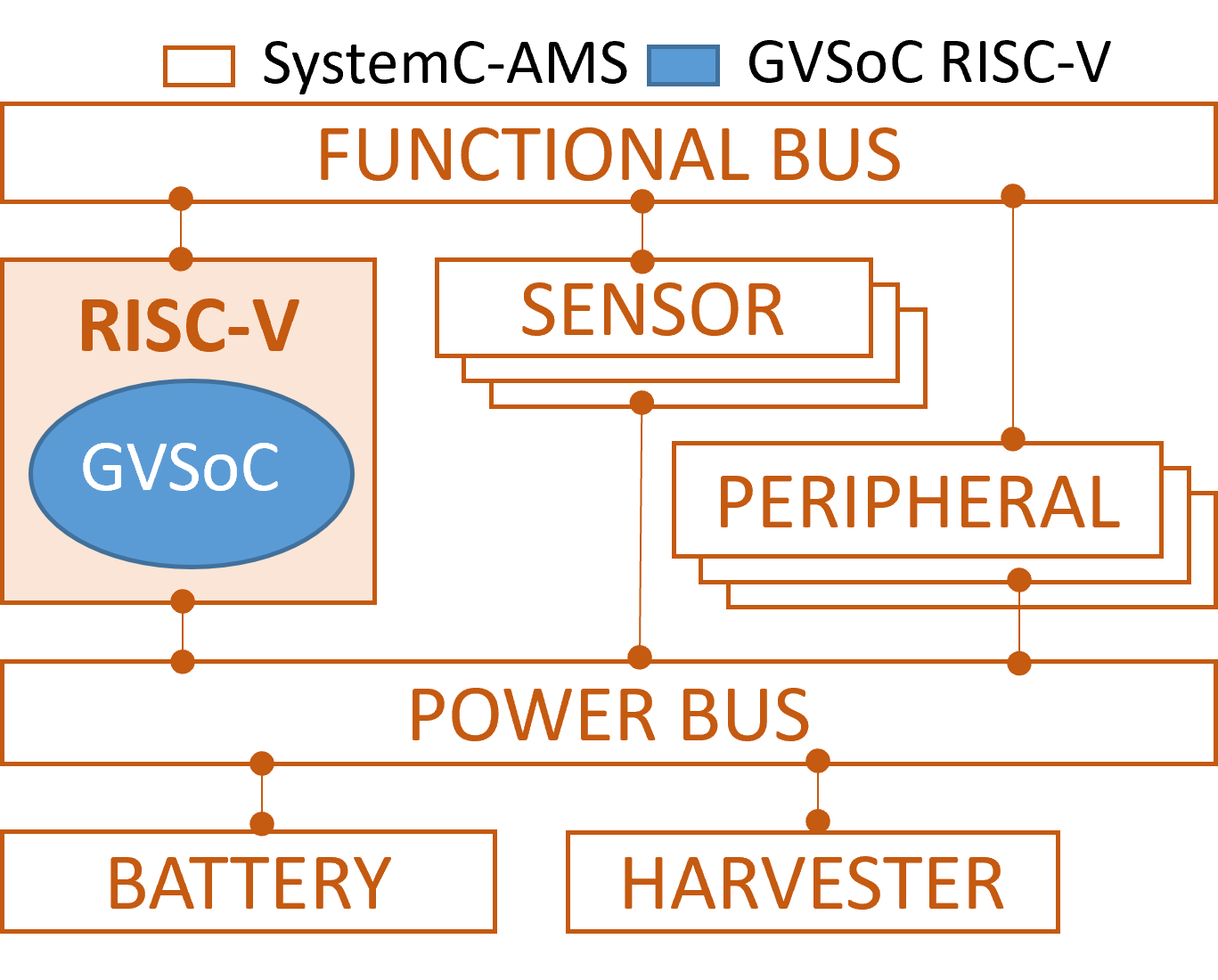}\vspace{-.2cm}
\caption{Bus-centric architecture adopted to simulate functional and power aspects of a RISC-V system.}\vspace{-.5cm}
\label{fig:highlevel}
\end{figure}

The coupling of functional and extra-functional simulation with SystemC-AMS has been first proposed in \cite{7812663}. The proposed architecture exploits a \emph{bus-centric paradigm} so that the simulation features one bus for each modeled aspect (in this case, functionality and power). Each system component may be connected to each bus by implementing one model per bus: e.g., a core will have a functional implementation that describes instruction processing connected to the functional bus and handles the timing behavior of the system, and a power model that estimates the corresponding power demand, exported to the power bus. %
Figure \ref{fig:highlevel} shows the overall architecture, where most system components are connected both to both buses. 

The mutual interdependency between the functional and the power models is achieved through an \emph{information exchange between the models of the same component}. In case of the core, the functional model may share its current state (e.g., the instruction being processed, active-idle-sleep state, etc) with the power model, which can estimate the corresponding voltage level and current demand. 

This architecture benefits from the modularity of SystemC-AMS, and allows to define the \emph{requirement that a RISC-V functional ISS must possess} to be compatible with our framework:
\begin{itemize}
    \item To ease the integration with SystemC-AMS, the simulator implementation must be C/C++-based; 
    \item The simulator must model instruction-level timing. This is fundamental to allow a precise estimation of power over time, accounting for the impact of the processor's execution;
    \item The simulator must be embeddable as a software component within a SystemC module and must expose APIs to export state information or direct power estimates. %
\end{itemize}
\begin{figure*}
    \centering
    \includegraphics[width=.8\linewidth]{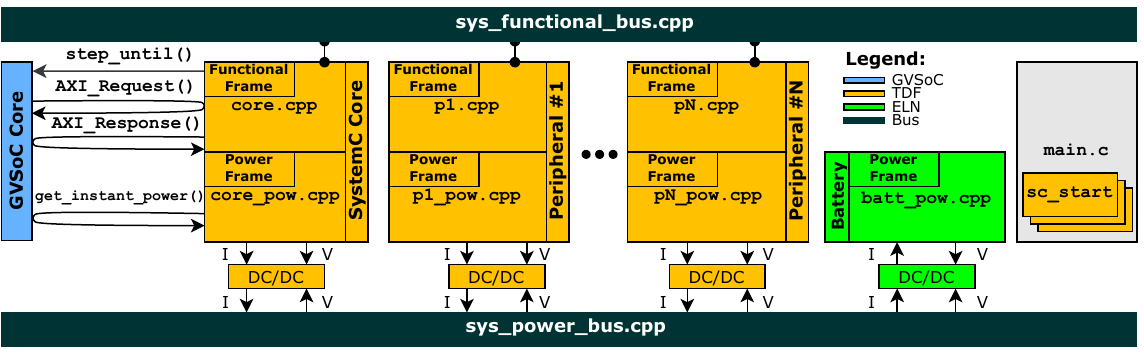}\vspace{-2mm}
    \caption{Implementation of the integration between SystemC-AMS and GVSoC.}\vspace{-.3cm}
    \label{fig:schema}
\end{figure*}
\subsection{The GVSoC RISC-V ISS}

GVSoC \cite{gvsoc} is an open-source C++ event-driven simulation platform for RISC-V cores of the Parallel Ultra-Low Power (PULP) family \cite{pulp}, supporting the modeling of low-power CPUs and the definition of complex full-platforms, including multicore, multi-memory levels (i.e., on- and off-chip), complex I/O peripherals, and accelerators \cite{gvsoc}. %
GVSoC supports virtual prototyping and DSE through early-stage performance evaluation based, e.g., on hardware counters and timing models, and it is \textit{almost cycle-accurate} (with a maximum error margin for cycle accuracy of 10\%). 

GVSoC simulation is event-driven. A circular buffer contains every event generated in the system, enqueued based on the corresponding latency. At any time, the simulator identifies the next event, processes the corresponding actions, and updates the queue. When not used as a stand-alone component, GVSoC exposes the {\tt step\_until()} API primitive, which runs the simulation until a specified time and returns the timestamp of the next event in the queue.

The latest version of GVSoC includes power models that account for both dynamic and leakage power of the different frequency domains and power islands within the modeled PULP chip. Power information can be retrieved by invoking the {\tt get\_instant\_power()} API primitive, which outputs the overall consumption of the system, aggregating all the contributions due to the set of events that occurred in the latest simulated timestamp.

In this work, we select GVSoC to demonstrate our SystemC integration, as it respects all requirements described at the end of Sec.~\ref{subsec:arch}. However, we remark that, in principle, our architecture is orthogonal to the specific RISC-V ISS.

%% file: sec/integration.tex
Fig.~\ref{fig:schema} shows the detailed implementation of the general simulation architecture discussed above. At the heart of the system is the SystemC core unit, which orchestrates the simulation and facilitates the interaction between SystemC and GVSoC. 
The simulation is organized as follows.
At the beginning, the {\tt sc\_main} function (implemented by the {\tt main.c} file) instantiates the GVSoC ISS and all SystemC-AMS components. Both GVSoC and SystemC initialize their event queues by starting SW execution (in the former case) and executing all components exactly once (in the latter case).

The subsequent simulation proceeds by alternating the execution of GVSoC and SystemC-AMS. Precisely: %
\begin{enumerate}
\item
The functional model of the core ({\tt core.cpp}) invokes the {\tt step\_until()} API function of GVSoC, which executes the SoC functionality and returns the timestamp $t$ of the following event in the GVSoC queue;

\item 
{\tt core.cpp} then executes a SystemC-AMS {\tt wait()} until $t$, to allow the execution of other SystemC-AMS components and the temporal alignment of the two simulators.

\item These two steps are repeated until the simulation is completed.
\end{enumerate}
Whenever the SW executing in GVSoC needs to communicate with a bus-connected functional component instantiated in SystemC-AMS (e.g., a peripheral), it propagates the necessary information (i.e., address, control signals, data) to the functional model of the core through an {\tt AXI\_Request()}, which is intercepted by a callback in {\tt core.cpp}. The functional model of the core then forwards this information to the SystemC functional bus, where it can be intercepted by the appropriate receiver (e.g., the peripheral's functional model).
Lastly, {\tt core.cpp} also collects from the functional bus any response received from peripherals. Such information is propagated to GVSoC with an {\tt AXI\_Response()} call, thus closing the cycle. 

Functional execution triggers the execution of the corresponding power models. In case of the core (i.e., {\tt core\_pow.cpp}), this corresponds to an invocation of the {\tt get\_instant\_power()} primitive of GVSoC. The power model of other peripherals is implemented directly in SystemC-AMS (i.e., {\tt *\_pow.cpp}). For instance, a peripheral might be modeled as a simple Power State Machine (PSM) sensitive to its operating mode (e.g., active-idle-sleep).

This dual-instance approach allows for a synchronized and detailed simulation of both the functional interactions and power consumption dynamics within the system.

%% file: sec/results.tex
We consider an audio processing task executed on a battery-operated device to evaluate the simulation methodology. We selected this use case since audio DSP is common in devices such as smart wearables, for which power simulation and optimization are critical. Furthermore, the PULP devices simulated by GVSoC have ISA extensions and architectural features specifically tailored for DSP applications~\cite{pulp}.
More specifically, we mimic an Adaptive Filtering workload, commonly found in applications such as noise cancellation, echo suppression, and dynamic equalization~\cite{fir-fpga}. To this end, we implement a simple software application in GVSoC that repeatedly stores input samples from a microphone into a buffer, processes them with a FIR filter with 40 taps, and then waits for the next buffer to be available before repeating the whole process.

As the main computing device, we use GVSoC's model for GAP9 \cite{gap9}, a PULP system comprising a cluster of 9 RISC-V cores to speed-up compute-intensive DSP workloads. We configure the SystemC-AMS power bus to deliver a regulated 3.3V supply to all components.
The rest of the simulation setup includes: i) a Mic Click microphone sensor from Mikroe~\cite{noauthor_mic_nodate},
whose power consumption, ranging in 120-160$\mu$A, is modeled as a state-sensitive PSM; ii) a Panasonic CG-425A 32mAh@3.8V battery~\cite{noauthor_cg-425am3_nodate}, simulated using the circuit-equivalent model of~\cite{model_type}, whose parameters are populated from datasheet information; iii) two DC-DC converters. While the microphone directly expects a 3.3V supply, the battery is connected to the power bus through a RT8097A converter~\cite{rt8097a}, and GAP9 is supplied at 1.8V by a ST1PS03~\cite{noauthor_st1ps03_nodate}. For both DC/DCs, we use LUT-based efficiency models that depend on their current output, which is extracted from datasheet curves.

\begin{figure}[t]
\centering
\includegraphics[width=\columnwidth]{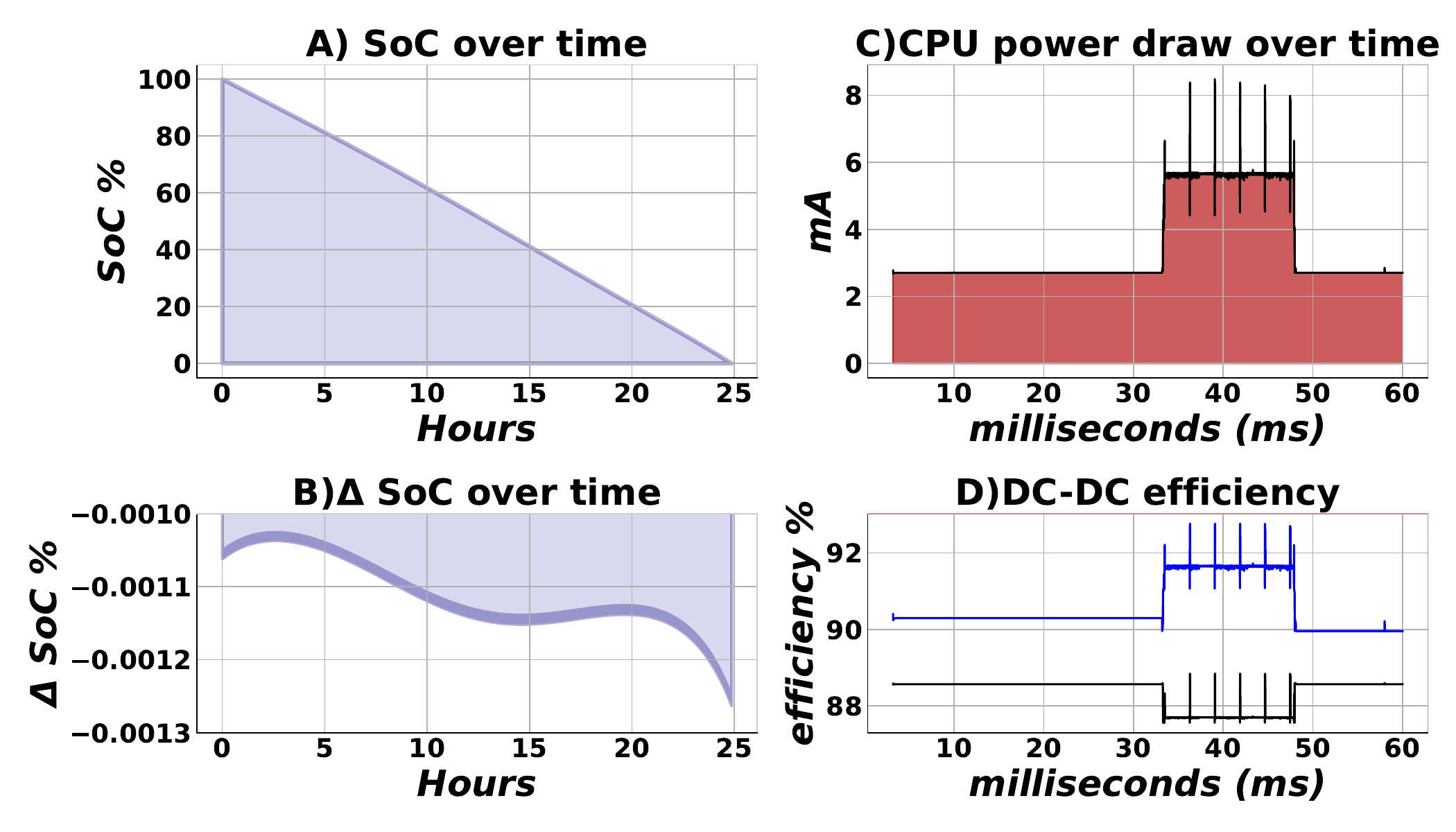}
\vspace{-.8cm}
\caption{Simulation of an audio filtering application.}
\label{fig;SoC}
\vspace{-.6cm}
\end{figure}

Fig.\ref{fig;SoC}-A shows the complete depletion of the battery throughout a simulation.
The battery can survive almost \textbf{25 hours} for this workload, showcasing the low-power nature of GAP9. The state of charge decreases almost linearly due to the low currents drawn by the sensor and computing subsystem. However, ``zooming in'' to look more in detail at the cumulated state-of-charge difference ($\Delta$SoC) in 1s intervals (Fig.~\ref{fig;SoC}-B), the non-linearity of the discharge behavior can be noticed. This is due to the variation of the battery's internal resistance embedded in our circuit-equivalent model.

Fig.~\ref{fig;SoC}-C displays the zoomed-in load power profile for a portion of an iteration in the simulated application. The plot clearly shows the phases that compose each iteration: i) the initial low and constant power consumption when GAP9 waits for new sensor data; ii) a region of higher consumption linked to the activation of GAP9's cluster, which also includes a set of power spikes, in correspondence to the start of inner processing loops in the cluster cores (the whole execution of the FIR is \textit{tiled} in smaller sections); iii) the final wait before the start of the following iteration. 

Fig.~\ref{fig;SoC}-D shows the efficiency of the DC/DC converters for the battery (blue curve) and core (black curve) over the same time period. This graph shows that the DC/DC selected for GAP has a lower efficiency overall, which, however, as desired, is higher during long idle phases and decreases during the short higher-power computation phases.

\subsection{Design Space Exploration}
In this section, we show an example of how our framework can be used for a DSE that considers both functional and power aspects. Namely, starting from the experiment of Fig.~\ref{fig;SoC}, which we denote as configuration \textbf{A}, we first reduce the processing workload by decreasing the filter taps to 20 (\textbf{B}). This mimics a quality-of-service reduction to increase the battery lifetime. Then, based on the results of Fig.~\ref{fig;SoC}, we try replacing the GAP9 DC/DC converter with the potentially more efficient RT8097A (\textbf{C}). Lastly, we combine both modifications (\textbf{D}). To stress the power differences, we reduce the interval between two filter executions from 1s, as in Fig.~\ref{fig;SoC}, to 50ms.

\begin{figure}[t]
    \begin{minipage}{.38\linewidth}
    \centering
    \includegraphics[width=\columnwidth]{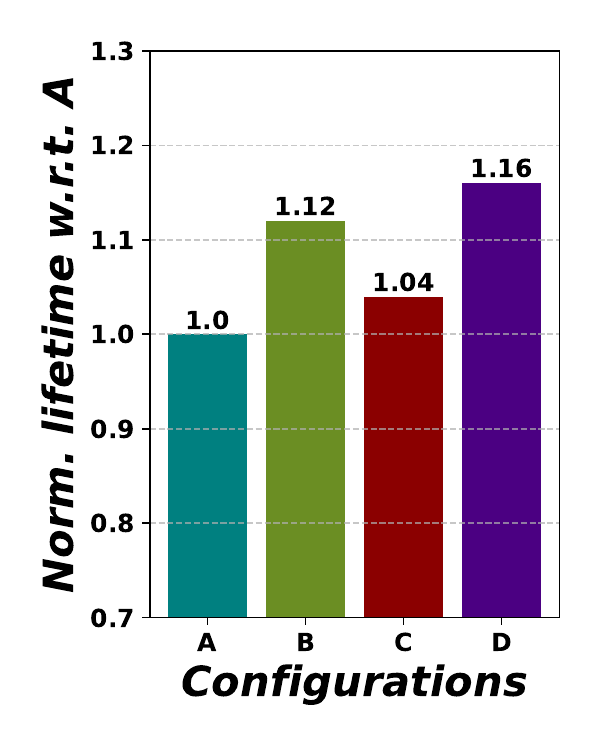}
    \vspace{-1cm}
    \caption{Battery life}
    \label{fig;dse}
    \end{minipage}
    \begin{minipage}{.6\linewidth}
        \footnotesize
        \renewcommand*{\arraystretch}{1.2}
        \centering
        \vspace{-0.3cm}
        \captionof{table}{Configurations comparison}
        \vspace{-0.3cm}
            \begin{tabular}{|p{0.6cm}|p{1cm}|p{1cm}|p{1cm}|}
                \hline
                \textbf{Conf.} & \textbf{Battery P. [mW]} & \textbf{DC/DC Eff. [\%]} & \textbf{$\Delta$SoC @1h [\%]}\\
                \hline
                A&  1.67& 88.3& 5.2\\
                \hline
                B&  1.49& 88.4& 4.6\\
                \hline
                C&  1.60& 91.6& 5.0\\
                \hline
                D&  1.43& 91.3& 4.4\\
                \hline
            \end{tabular}
        \label{tab;dse}
    \end{minipage}
    \vspace{-0.5cm}
\end{figure}

The results of this exploration are shown in Table~\ref{tab;dse}, which reports the average power drawn from the battery (Battery P.), the average GAP9 converter efficiency (DC/DC Eff.), and the $\Delta$SoC over 1 hour of simulation. Fig~\ref{fig;dse} shows the estimated battery lifetime of the four configurations, normalized to the one of configuration \textbf{A}.

As shown, reducing the GAP9 workload (\textbf{B}) leads to an estimated 12\% increase in battery life, mainly due to lower power demand and secondarily to a slightly higher DC/DC efficiency (which is higher when GAP9 is not computing, as shown in Fig.~\ref{fig;SoC}). Conversely, replacing the DC/DC only prolongs the battery life by 4\%, but without impacting quality-of-service (\textbf{C}). Lastly, combining the two optimizations (\textbf{D}) leads to a 16\% lifespan improvement.

Overall, the results of Fig.~\ref{fig;SoC}-\ref{fig;dse} and Table~\ref{tab;dse} showcase the type of fine-grained power analysis enabled by the proposed simulation framework. These insights are obtained at a limited cost in terms of simulation time overhead with respect to vanilla GVSoC, i.e., approximately 7.3\% when GVSoC's power model is active (for a purely functional simulation, disabling GVSoC's power model, the overhead would grow to approximately 50\%).

%% file: sec/conclusions.tex
We have introduced a fully open-source framework to simulate functional and non-functional properties of RISC-V systems, with a specific focus on power consumption.
On a realistic audio processing use case, we have shown how our framework can be leveraged for DSE and battery lifespan estimation.
Our proposed architecture is fully modular, and can support much more complex configurations (e.g. higher n. of components) than the one shown here. Moreover, all models can be swapped with alternative implementations depending on the desired trade-off between accuracy and simulation speed, including functional and power models of the computing system and peripherals, and power models of the battery and DC/DC converters.
Our future works include the extension of the framework to support other non-functional properties (e.g., thermal behaviour), and its application to more complex use cases.